%% file: neurips.tex
\documentclass{article}

\usepackage[final,nonatbib]{neurips_2022}

\usepackage[utf8]{inputenc} %
\usepackage[T1]{fontenc}    %
\usepackage{hyperref}       %
\usepackage{url}            %
\usepackage{booktabs}       %
\usepackage{amsfonts}       %
\usepackage{nicefrac}       %
\usepackage{microtype}      %
\usepackage{xcolor}         %
\usepackage{tikz}
\usepackage{comment}
\usepackage{amsmath,amssymb} %
\usepackage{color}
\usepackage{makecell}
\usepackage{array,multirow,graphicx}
\usepackage{subcaption}
\usepackage{lscape} 
\usepackage{lmodern}
\usepackage{todonotes}

\usepackage{enumitem}
\setlist[itemize]{itemsep=0pt,topsep=-3pt,partopsep=0pt,parsep=0pt,leftmargin=4mm}
\newcommand{\mappingTwoToSix}{\phi}

\newcommand{\pdffunction}{\mathcal{P}}

\newcommand{\shape}{\mathcal{S}}
\newcommand{\mapTwoToThree}{\phi}
\newcommand{\mapSixToTwo}{\psi}
\newcommand{\mapTwoToSix}{\varphi}
\newcommand{\proba}{\mathcal{P}}
\newcommand{\pointtwod}{\mathbf{x_{\mathcal{P}}}}
\newcommand{\pointthreed}{\mathbf{x_{\mathcal{S}}}}
\newcommand{\pointgeneral}{\mathbf{x}}

\newcommand{\featurepsii}{\mathbf{f}_{\psi,i}}
\newcommand{\featurephii}{\mathbf{f}_{\varphi,i}}
\newcommand{\featurepsi}{\mathbf{f_{\psi}}}
\newcommand{\featurephi}{\mathbf{f_{\varphi}}}

\usepackage{colortbl}

\definecolor{yellow}{rgb}{1,1, 0.6}
\definecolor{lightyellow}{rgb}{1,1, 0.8}
\definecolor{orange}{rgb}{1, 0.8, 0.6}
\definecolor{tabred}{rgb}{1, 0.6, 0.6}

\title{Learning Joint Surface Atlases}

\author{%

  \textbf{Theo Deprelle}$^{1*}$ \\
   \And
   \textbf{Thibault Groueix}$^2$ \\
   
   \AND
   \textbf{Noam Aigerman}$^2$ \\
   
   \And
   \textbf{Vladimir G. Kim}$^2$ \\

   \And
   \textbf{Mathieu Aubry}$^1$ \\
  \And
  {\normalfont$^1$LIGM, Ecole des Ponts, Univ Gustave Eiffel, CNRS, $^2$Adobe Research}
}

\begin{document}

\maketitle
\begin{abstract}

This paper describes new techniques for learning atlas-like representations of 3D surfaces, i.e. homeomorphic transformations from a 2D domain to surfaces. %
Compared to prior work, %
we propose two major contributions. First, instead of mapping a fixed 2D domain, such as a set of square patches, to the surface, we learn a continuous 2D domain with arbitrary topology by optimizing a point sampling distribution represented as a mixture of Gaussians. Second, we learn consistent mappings in both directions: charts, from the 3D surface to 2D domain, and parametrizations, their inverse. 
We demonstrate that this improves the quality of the learned surface representation, as well as its consistency in a collection of related shapes. It thus leads to improvements for applications such as correspondence estimation, texture transfer, and consistent UV mapping. 
As an additional technical contribution, we outline that, while incorporating normal consistency has clear benefits, it leads to issues in the optimization, and that these issues can be mitigated using a simple repulsive regularization. We demonstrate that our contributions provide better surface representation than existing baselines. 
\end{abstract}

\begin{figure}[!h]
    \includegraphics[width=\textwidth]{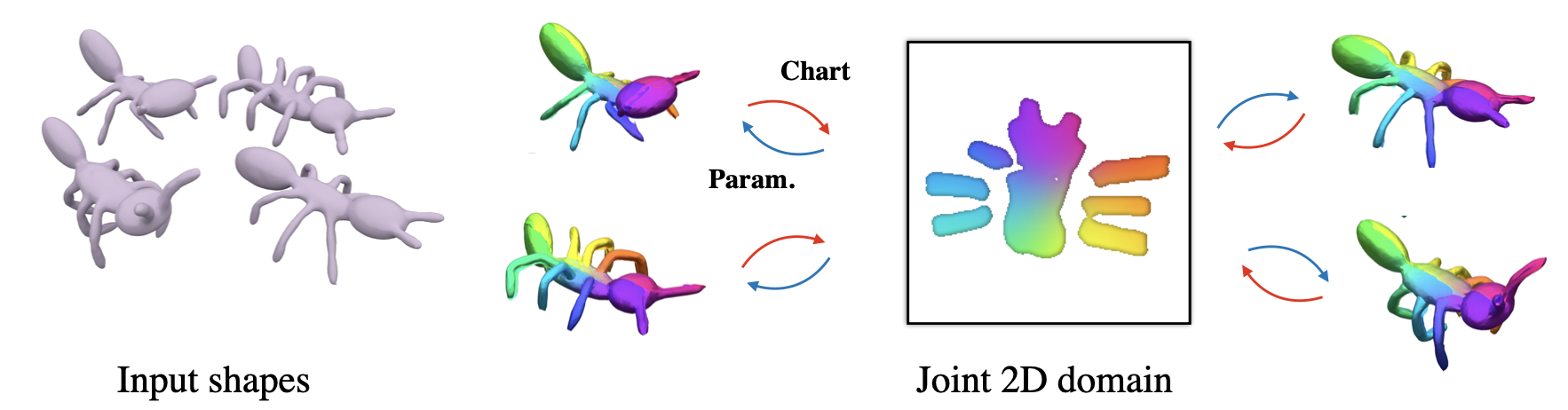}
    \centering
    \caption{
    From a collection of shapes without annotations, we learn a 2D domain which can be used to parameterize all shapes, parametrizations (blue) and chart-mappings (red).}
    \label{fig:teaser}
    \vspace{-1em}
\end{figure}
\section{Introduction}
\input{content/intro}

\section{Related work}
\input{content/related_work}

\input{content/method}

\section{Experiments}

\input{content/experiments}

\vspace*{-5pt}
\section{Conclusion}
\vspace*{-5pt}

\input{content/conclusion}

\textbf{Acknowledgments} Thanks to F. Darmon, R. Loiseau, E. Vincent for their feedbacks on the manuscript; E. Shechtman
, D. Picard, Y. Siglidis and G. Ponimatkin for inspiring discussions; and B. George for code suggestions; This work was partly supported by ANR project EnHerit ANR-17-CE23-0008, Labex Bézout, gifts from Adobe to École des Ponts and HPC resources from GENCI-IDRIS (2021-AD011011937R1). 
\bibliographystyle{splncs04}
\bibliography{egbib}

\end{document}

%% file: content/intro.tex
This paper is concerned with 3D surfaces and their representation as atlases or UV maps, i.e., their mappings to and from a domain of the 2D Euclidean space. Surfaces and altases are closely related: surfaces are $2$-manifolds embedded in $\mathbb{R}^3$ and are defined in topology by the existence of \emph{charts}, homeomorphic mappings from the surface to a 2D Euclidian domain. This relation is also central to many algorithms related to surfaces: on one hand, the computation of UV maps of meshes %
is a highly-active research topic:  %
on the other hand, deep learning works have successfully used atlas-like representation and learn local parametrizations to represent 3D surfaces~\cite{bednarik2020shape,groueix2018}. These last techniques %
typically learn to map a fixed sets of 2D squares to 3D, which can approximate a 3D surface and makes it possible to use the 2D domain to compute correspondences between \emph{predicted} surfaces~\cite{bednarik2021temporally} for which parametrizations are learned jointly.

While these parametrization-based methods produce rather-accurate 3D surface reconstructions, %
they do not lead to a well-defined homeomorphic map between a 2D domain and the predicted surface, which limits the scope of applications of these techniques. For examples, when mapping a set of 2D squares to 3D, AtlasNet~\cite{groueix2018} leads to many overlaps, and the chart-mappings from 3D to 2D are thus not well defined. The predicted maps from 2D to 3D might also include large amount of distortion, thus not yielding a good UV map of the generated 3D surface, and for example preventing using it to define mappings between different input surfaces. Previous works have attempted to address these limitations in various ways. Williams et al.~\cite{williams2019deep} performs optimization for a single shape on local neighbourhoods with Earth Mover Distance to obtain homeomorphic parametrizations and consistent transitions maps, but this leads to an heavy optimization, with many local parametrizations, and has no obvious extension to multiple shapes. Rather than enforcing consistency between the mappings of square patches AtlasNet v2~\cite{deprelle2019learning} attempts to learn the 2D domain, e.g., by learning the positions of a fixed set of points, but loses the continuous aspect of the mapping and can still lead to overlapping patches. DSR~\cite{bednarik2020shape} keeps the AtlasNet framework and its intrinsic limitations, but uses several regularizations to encourage conformal mappings, to minimize the 3D overlap between the images of the square patches and to prevent patch collapse. However, it is still limited to square patches and we found it difficult to use on complex shapes. %

We argue that the two tasks of parameterizations and charts prediction are complementary to one another, and that learning the 3D shape(s) reconstruction should go together with learning a relevant 2D domain. We present an architecture for such a joint optimization, where we learn the 2D domain by learning a 2D probability distribution defined as a mixture of Gaussians and from which we sample points for reconstructing the surface.
These sampled points are mapped to 3D and compared (via chamfer distance) to the target point cloud; similarly, the point cloud is mapped to 2D and compared to the sampled points. We optimize for cyclic consistency between the two mappings, as well as for geometric losses such as isometric regularization. %
Through experiments, we show that our method is able to better reconstruct surfaces than existing baselines, in particular leading to more meaningful parametrizations with fewer artefacts and yielding meaningful correspondences between shapes in a collection. Our code is available on our \href{https://imagine.enpc.fr/~deprellt/joint-surface/}{project webpage}\footnote{\href{https://imagine.enpc.fr/~deprellt/joint-surface/}{https://imagine.enpc.fr/~deprellt/joint-surface/}}

%% file: content/related_work.tex
Our work is related to prior work in optimizing chart-mappings for UV parametrization and learning parameterization for surface reconstruction.

\paragraph{Optimizing charts-mappings for UV parametrization.}
Identifying chart is a long-standing problem in geometry processing~\cite{Sheffer:meshparam:2007}. Most prior techniques take a 2-manifold input represented as a mesh and map each point to a 2D domain. These methods typically aim to produce a bijective mapping~\cite{Smith:bijective:2015}, while also minimizing some distortion metric such as Dirichlet~\cite{Pinkall93computingdiscrete}, ARAP~\cite{liu2008local}, LSCM~\cite{levy2002lscm}, and symmetric Dirichlet~\cite{Rabinovich:SLIM:2017}.
Some techniques also aim to predict consistent chart-mappings for a collection of related shapes, so that semantically-similar points on different meshes map to the same point in 2D. Doing so enables many applications such as correspondence estimation~\cite{Aigerman2015}, morphing~\cite{Kraevoy2003}, and texture transfer~\cite{praun2001consistent}. Unlike our method, these techniques do not use deep learning and typically require manual input, such as a sparse set of corresponding points. By using neural networks to represent the UV map we can learn consistent charts without the user input or explicit supervision. %
We can also co-parameterize point clouds without knowing the underlying mesh. To the best of our knowledge, our work is the first method to use neural networks to learn consistent chart-mappings.

\clearpage
\paragraph{Learning surface parameterization for reconstruction.}
Previous learning-based techniques for surface parameterization mostly focused on modeling 2D to 3D map for reconstruction. To learn such a mapping from a collection of shape, AtlasNet~\cite{groueix2018} and FoldingNet~\cite{Yang18} pioneered the idea of using a Multi-Layered Perceptrons (MLP) as a class of continuous parametric function that embed a 2-manifold into 3D.  By conditionning the MLP on a latent code, and using a large-scale 3D shape repository~\cite{shapenet2015}, AtlasNet~\cite{groueix2018} demonstrated the possibility to predict the surface parameterization from an input mesh, point-cloud or even a single image. Perhaps surprisingly,  Williams et al.~\cite{williams2019deep} showed that parameterization did not necessarily need the regularization brought by learning on a collection of shapes, and could also be optimized on individual shapes.

Several approaches introduced novel losses to improve the parameterization. Deng et al.~\cite{deng2020better} improve the global arrangement of the different parts of the parameterization via a normal-aware reconstruction loss and a stiching loss. In DSR, Bednarik et al.~\cite{bednarik2020shape} regularize the smoothness of the reconstructed surface by optimizing a conformal energy, based on the Jacobians of the mappings. Remarkably, having access to analytical jacobians in a higher-order differentiation procedure opens-up exciting applications. For instance, Bednarik et al.~\cite{bednarik2021temporally}  achieve  temporally coherent parameterizations for each frame in a video by regularizing the deformation to locally have a constant metric tensor. 

Particularly relevant to us is AtlasNet-v2~\cite{deprelle2019learning} which jointly optimize the shape of the 2D manifold and learns the surface parameterization using two proposed strategies.  
The first one consists in learning a deformation of a fixed template into an elementary structure common to all shapes via an additional AtlasNet-like MLP . %
This strategy leads to elementary structures that can easily be meshed, but have the same topology as the initial template. %
The second strategy consists in sampling a fixed set of points on the template and adding them to the optimizer. After optimization, the sampled points can form a complex elementary structure, without topology constraints,  %
but meshing it is not straight-forward. 
In this work, we propose a method to combine the best features of both strategies, namely the ability to learn topologically complex structures and the ability to mesh them. %
We achieve this via a novel differentiable layer to sample points from 2D gaussians with a learnable mean (see Sec.~\ref{subsec:sampling}).

Our approach is different from AtlasNet~\cite{groueix2018} and its variants~\cite{badki2020meshlet,bednarik2021temporally,bednarik2020shape,deng2020better,deprelle2019learning,pang2021tearingnet,williams2019deep}  in that: (i) we aim to jointly learn the surface parametrizations and their inverse functions, the charts-mappings; (ii) we learn a 2D domain relevant to a family of shapes by optimizing a probability density function in the 2D Euclidean plane. We also differ in two novel losses that correctly orient the normals of the reconstructed surface and fix point-collapse in 3D which is a common artifact of AtlasNet-type of approaches.

%% file: content/method.tex
\begin{figure}[t!]
    \centering
    \includegraphics[width=\linewidth]{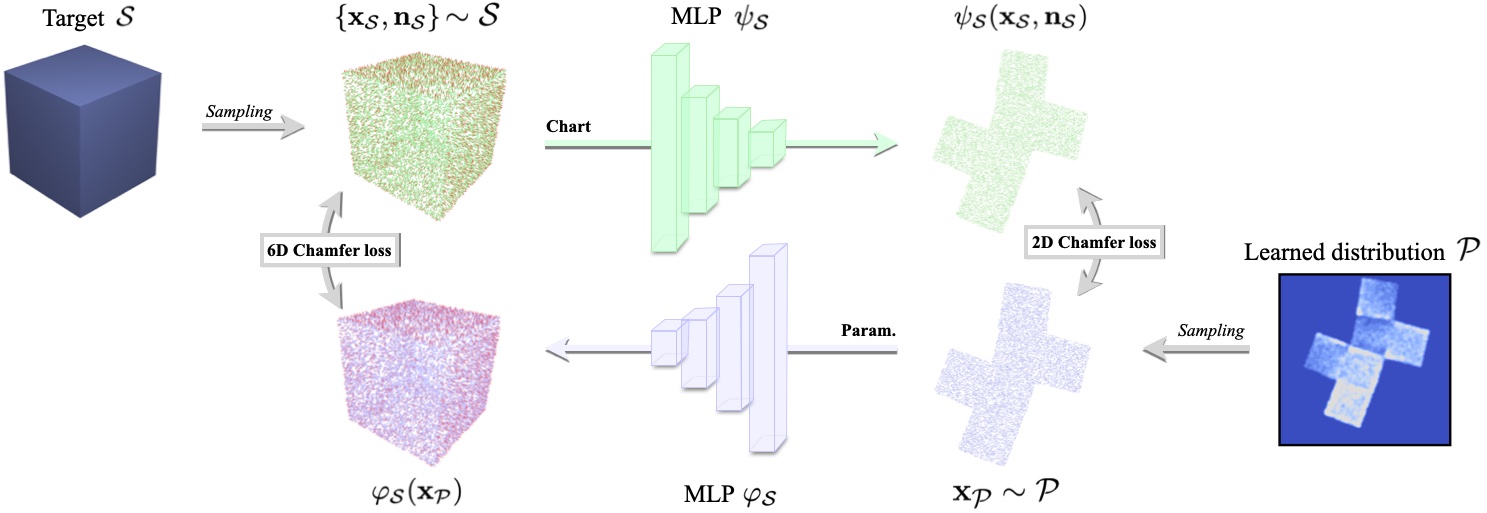}%

    \caption{\textbf{Single shape architecture.} %
    Given a shape $\shape$, we optimize jointly a 2D probability distribution $\proba$, a parametrization network, $\mapTwoToSix_\shape$ and chart-mapping network $\mapSixToTwo_\shape$. 
    The parameterisation network $\mapTwoToSix_\shape$ takes as input a point $ \pointtwod \in \mathbb{R}^2$ randomly sampled according to the density probability function $\proba$ to produce the  6D vector (3D point and normal coordinates) $\mapTwoToSix_\shape(\pointtwod)$. 
    The chart-mapping network $\mapSixToTwo_\shape$ takes as input a point $\pointthreed$ sampled on the shape $\shape$, its normal $\mathbf{n_\shape}$ and outputs a 2D points $ \mapSixToTwo_\shape(\pointthreed, \mathbf{n_\shape})$. 
    }
    \label{fig:archi}
    \vspace{-2em}
\end{figure}

\section{Method}

\paragraph{Overview.} Given a collection of shapes, our goal is to learn for all shapes surface parameterizations %
with their inverse chart-mappings and a join 2D domain on which the parametrizations are defined. 
For simplicity, we first present our approach in the case of a single shape $\shape$: in Section~\ref{subsec:overview} we explain how to model surface parameterization and chart-mapping, and introduce our main architectural blocks; in Section~\ref{subsec:sampling} we explain how we learn the 2D UV domain as a probability distribution; in section~\ref{subsec:losses} we discuss our losses. Our pipeline for a single shape is illustrated in Figure~\ref{fig:archi}. Finally, in Section~\ref{subsec:family} we explain how to train our approach jointly on a family of shapes.

\paragraph{Notations.} We use the following notations:
\begin{itemize}
    \item  $\shape$ : 3D shape of interest
    \item  $\proba$ : learned probability distribution in 2D%
    \item  $\mapTwoToThree_\shape : \mathbb{R}^2 \xrightarrow[]{} \mathbb{R}^3$ : surface parameterization 
    \item  $\mapTwoToSix_\shape : \mathbb{R}^2 \xrightarrow[]{} \mathbb{R}^6$ : surface parameterization with normals
    \item  $\mapSixToTwo_\shape : \mathbb{R}^6 \xrightarrow[]{} \mathbb{R}^2$ : chart-mapping
\end{itemize}
 We slightly abuse the brackets notation to indicate sampling a set of $M$ points, e.g.: %
\begin{itemize}
    \item  $\{\pointthreed\}$ is a set of $M$ points sampled on $\shape$ using a uniform probability distribution
    \item  $\{\pointtwod\}$ is a set of $M$ points sampled in $\mathbb{R}^2$ according the probability distribution $\proba$ 
\end{itemize}
\subsection{Parametrization and chart-mapping}
\label{subsec:overview}
We explain the two main components of our architecture for a given shape $\shape$, a chart-mapping $\mapSixToTwo_\shape$  and a surface parameterization network $\mapTwoToSix_\shape$.

\paragraph{Chart-mapping.} We %
learn a single chart-mapping from the shape $\shape$ to $\mathbb{R}^2$, using a Multi-Layer Perceptron (MLP) with both point coordinates and normals as input. We observe that naively learning an $\mathbb{R}^3$ to $\mathbb{R}^2$ mapping leads to the collapse of thin surfaces after their mapping to the 2D domain. Indeed, coordinate-based MLPs are continuous functions and Tancik et. al.~\cite{tancik2020fourier} showed that they have a prior to learn smooth functions in the absence of positional encoding. Hence, two 3D points with very close spatial coordinates but opposite normals tend to be mapped closely in 2D. For chart-mappings, such smoothness is generally a desirable feature that should be maintained, but distant normals should be a strong cue to indicate that points are intrinsically far. To handle thin surfaces, we propose to learn a mapping $\mapSixToTwo_\shape$ from  $\mathbb{R}^6$ to $\mathbb{R}^2$ that takes as input a point  $\pointthreed$ and its associated normal  $\mathbf{n_{\shape}}$ scaled to have norm $\alpha$. The parameter $\alpha$ is a hyperparameter of our approach which controls how much normals contribute to distances compared to the 3D positions of the points.We set it to $0.01$ in all our experiments.  %

\paragraph{Surface parameterization.} For a shape $\shape$, we seek to learn the inverse function of the chart-mapping, $\mapSixToTwo_\shape^{-1} : \mathbb{R}^2 \xrightarrow[]{} \mathbb{R}^6$. We first use an MLP $\mapTwoToThree_\shape : \mathbb{R}^2 \xrightarrow[]{} \mathbb{R}^3$ to parameterize the surface $\shape$.  The function $\mapTwoToThree_\shape$ maps a point $\pointgeneral=(u,v)$ in $\mathbb{R}^2$ into a point $\mapTwoToThree_\shape(\pointgeneral)$ in $\mathbb{R}^3$. %
The Jacobian $\mathbf{J}_\pointgeneral$ of the mapping $\mapTwoToThree_\shape$ is defined at every point $\pointgeneral$ by:%
\begin{equation}
   \label{equation:normal1}
   \mathbf{J}_{\pointgeneral}%
   = 
   \big[
   \mathbf{J}_{\pointgeneral,u}
   ,
   \mathbf{J}_{\pointgeneral,v}
   \big] 
   = 
   \big[
   \frac{\partial\mapTwoToThree_\shape}{\partial \mathit{u}}(\pointgeneral)
   ,
   \frac{\partial\mapTwoToThree_\shape}{\partial \mathit{v}}(\pointgeneral)
   \big] ~.
\end{equation} 
The two partial derivatives are trivial to compute with Pytorch auto-differentiation~\cite{paszke2019pytorch}. The normal $\mathbf{n}$ to the parametrizedsurface at  $\mapTwoToThree_\shape(\pointgeneral)$ can be computed as the normalised cross product of $\mathbf{J}_{\pointgeneral,u}$ and $\mathbf{J}_{\pointgeneral,v}$ and scaled by the same hyperparameter $\alpha$ used for the chart-mapping $\mapSixToTwo_\shape$:
\begin{equation}
     \label{equation:normal2}
      \mathbf{n}
     = \alpha \frac{\mathbf{J_{\pointgeneral,\mathit{u}}} \times \mathbf{J_{\pointgeneral,\mathit{v}}}}{\|\mathbf{J_{\pointgeneral,\mathit{u}}} \times \mathbf{J_{\pointgeneral,\mathit{v}}}\|}.
\end{equation}
 We %
 define $\mapTwoToSix_\shape : \mathbb{R}^2 \xrightarrow[]{} \mathbb{R}^6$ as the concatenation of an output point and its scaled normal:
\begin{equation}
    \mapTwoToSix_\shape(\pointgeneral) = [\mapTwoToThree(\pointgeneral), \mathbf{n}]~.
\end{equation}
To summarize, the mapping $\mapTwoToSix_\shape$ is designed to represent the inverse of the chart-mapping $\mapSixToTwo_\shape$. However, it is not defined for every point in $\mathbb{R}^2$, and in the next section we focus on identifying the 2D domain for which it is defined.

\subsection{Learning a 2D domain as a sampling probability distribution}
\label{subsec:sampling}

To parametrize a shape $\shape$, we want to define a domain in $\mathbb{R}^2$ such that $\mapTwoToThree_\shape$ defines a bijection from this 2D domain to $\shape$. In practice, during training, we want to learn this domain, sample points inside it, and map them using $\mapTwoToSix_\shape$. 
Instead of handling explicitly the 2D domain geometry, e.g., points or primitives similar to~\cite{deprelle2019learning}, we take a probabilistic approach and learn the parameters of a probability distribution $\proba$ from which to sample points. This enables us to easily deal with topological changes. We now detail how we represent this probability distribution, learn it, and use it to define a 2D domain. %

\paragraph{Modelling 2D sampling probability as a mixture of Gaussians.} We sample 2D points according to a probability distribution $\proba$ which we model as a mixture of $K$ 2D Gaussians 
with means $\mathbf{\mu}_i \in \mathbb{R}^2$ for $i=1, ... , K$ , a fixed standard deviation $\mathbf{\sigma} \in \mathbb{R}$ and fixed mixing coefficients equal to $1/K$:
\begin{equation}
    \proba(\pointgeneral) =  \frac{1}{K} \sum_{k=1}^K  \mathcal{N}(\pointgeneral | \mathbf{\mu}_i, \mathbf{\sigma}) \quad \text{with} \quad \sigma = \frac{1}{\sqrt{K}}~.
    \label{eq:gaussian}
\end{equation}
During training, at every iteration, we sample $N$ 2D points $\{ \pointtwod \}$ from   $\proba$, which are both the input of the parameterisation network $\mapTwoToSix_\shape$ and the target of the chart-mapping network $\mapSixToTwo_\shape$.

\paragraph{Learning the Gaussian means.} To allow learning the 2D domain, we make the  means $\mu_i$ learnable parameters of the method. To do so, we need to see the sampled points as differentiable with respect to the $\mu_i$. We achieve this with the pathwise gradient estimator from \cite{kingma2013auto}, also called the reparameterization trick. This consist in expressing a parameterized random variable via a parameterized deterministic function of a parameter-free random variable. For Gaussian Mixture Models (GMM), this simply amounts to sampling a GMM with zero means and adding the means to the sampled points, i.e., defining each point $\pointtwod$ as the result of the following process: first selecting the id $i$ of a mixture component using a uniform distribution; then sampling a 2D point $\pointgeneral_\mathcal{N}$ from a Gaussian distribution of mean $0$ and standard deviation sigma $\pointgeneral_\mathcal{N} \sim \mathcal{N}(0, \mathbf{\sigma})$; finally, defining the point $\pointtwod$ as $ \pointtwod = \pointgeneral_\mathcal{N} + \mathbf{\mu}_i$, which is trivially differentiable with respect to $\mathbf{\mu}_i$.
Please see Figure~\ref{fig:repulsion} for examples of learned probability distributions.

\paragraph{From probability distribution to 2D continuous domain.} Once the distribution has been optimized, we simply threshold the probability distribution function $\proba$ to obtain a 2D domain. We can then compute a 2D triangulation of the domain, and use the parameterization network $\mapTwoToThree$ to obtain a 3D mesh %
(as can be seen in Figure~\ref{fig:collection}).

\subsection{Training losses}
\label{subsec:losses}
We now explain the loss function we optimize.  %
We write our loss for a single shape:
\begin{equation}
        \mathcal{L}_{single}(\mapTwoToSix_\shape , \mapSixToTwo_\shape, \mathbf{\mu}) =       
        \lambda_{\textrm{6D}}  \mathcal{L}_{\textrm{6D}} +  
        \lambda_{\textrm{2D}} \mathcal{L}_{\textrm{2D}} + 
        \lambda_{\textrm{cycle}} \mathcal{L}_{\textrm{cycle}} + 
        \lambda_{\textrm{iso}}  \mathcal{L}_{\textrm{iso}} +
        \lambda_{\textrm{rep}} \mathcal{L}_{\textrm{rep}} ~,
\end{equation}
where $\mu=(\mu_1, ... , \mu_K)$, the $\lambda$ are scalar hyper-parameters and the different loss terms are detailed bellow. To compute distances between two sets of points $\mathcal{X}$ and $\mathcal{Y}$, we base our losses on the Chamfer distance defined as:
\begin{equation}
        \mathcal{L}_\textrm{chamfer}(\mathcal{X}, \mathcal{Y})
        = %
        \frac{1}{|\mathcal{X}|}
        \sum_{\mathbf{x}\in\mathcal{X}}
        \min_{\mathbf{y}\in\mathcal{Y}} 
         \|\mathbf{x} - \mathbf{y}\|^2  +  %
        \frac{1}{|\mathcal{Y}|}
        \sum_{\mathbf{y}\in\mathcal{Y}}
        \min_{\mathbf{x}\in\mathcal{X}} 
        \|\mathbf{y} - \mathbf{x}\|^2 ~,
\end{equation}
where $|\mathcal{X}|$ and $|\mathcal{Y}|$ are the number of points in $\mathcal{X}$ and $\mathcal{Y}$ respectively.
Our set of losses is designed to enforce two objectives : (i) ensuring an accurate surface parameterization $\mapTwoToSix_\shape$ with low distortion, (ii) ensuring that $\mapTwoToSix_\shape$ and $\mapSixToTwo_\shape$ are indeed inverse of each other.

\paragraph{Surface parameterization reconstruction loss.}
The surface parameterization $\mapTwoToSix_\shape$ %
takes as input a set of 2D points $\{ \pointtwod \}$ sampled from the probability distribution $\proba$ %
and outputs a set of 6D points (3D points with scaled normals). 
We minimize the 6D Chamfer distance between the set of generated points  $\{ \mapTwoToSix_\shape(\pointtwod) \}$ and a set of points $\{ \pointthreed \}$ associated to normals $\{\mathbf{n_\shape} \}$ sampled on the target shape $\shape$:
\begin{equation}
        \mathcal{L}_{\textrm{6D}}(\mapTwoToSix, \mathbf{\mu})
        = \mathcal{L}_\textrm{chamfer}(\{ \mapTwoToSix_\shape(\pointtwod) \}, \{\pointthreed,\mathbf{n_\shape}\})~.
\end{equation}

\paragraph{Chart-mapping reconstruction loss.}

We encourage the overall 2D projection of the shape under $\mapSixToTwo_\shape$ and the probability distribution $\proba$ to be the same.  Recall that $\mapSixToTwo_\shape$ takes as input
$\{ \pointthreed, \mathbf{n_\shape} \} $ (3D points uniformely sanpled on $\shape$ with scaled normals) and outputs a set of 2D points $ \{\mapSixToTwo_\shape(\pointthreed,\mathbf{n_\shape})\}$.
We simply minimize the Chamfer distance between the generated 2D points and a set of points $\{ \pointtwod \}$ sampled from $\mathcal{P}$:

\begin{equation}
        \mathcal{L}_{\textrm{2D}}(\mapSixToTwo, \mathbf{\mu})
        = \mathcal{L}_\textrm{chamfer}(\{\mapSixToTwo_\shape(\pointthreed,\mathbf{n_\shape })\} , \{\pointtwod\})~.
\end{equation}

\paragraph{Cycle-consistency loss.}

We want the two mappings $\mapTwoToSix$ and $\mapSixToTwo$ to be inverse of one another. We encourage this using a cycle-consistency loss on 2D points sampled from $\proba$ and on 3D points sampled uniformly on $\shape$ with their associated normals:

\begin{equation}
    \mathcal{L}_{\scriptsize\textrm{cycle}}(\mapTwoToSix, \mapSixToTwo, \mathbf{\mu})
    =   \frac{1}{M}  \sum_{\pointgeneral\in\{\pointtwod\} } 
    \|\pointgeneral - \mapSixToTwo \circ \mapTwoToSix(\pointgeneral) \|^2 
    + 
    \frac{1}{M} \sum_{(\pointgeneral,\mathbf{n})\sim\{\pointthreed,\mathbf{n_\shape}\} } 
    \|\pointgeneral - \mapTwoToSix \circ \mapSixToTwo(\pointgeneral,\mathbf{n}) \|^2 ~.
\end{equation}

\paragraph{Distortion regularization loss.}
We limit distortion in the parameterization with an isometric regularization.
Given the Jacobian $\mathbf{J}_\pointgeneral$ of the transformation $\mapTwoToThree$ at point $\pointgeneral$ (Equation ~\ref{equation:normal1}) and $\mathbf{I}$ the identity matrix,
the isometric loss can be written:
\begin{equation}
    \label{eq:iso}
    \mathcal{L}_{\scriptsize\textrm{iso}}(\mapTwoToSix_\shape, \mathbf{\mu}) = \frac{1}{M} \sum_{\pointgeneral\in\{\pointtwod\} }  \|\mathbf{J}_\pointgeneral\mathbf{J}_\pointgeneral^T - \mathbf{I} \| ~,
\end{equation}
where the sum is over points sampled according to $\proba$. 
As already observed in \cite{bednarik2021temporally,groueix2018b}, this type of regularization has the additional benefit of making the parameterizations more consistent across shapes. 

\paragraph{Probability distribution regularization loss.}
Non-uniform density is a known failure mode of the Chamfer distance, as shown in Figure~\ref{fig:qualitative_ablation} and also observed in~\cite{bednarik2020shape}. In theory,  a loss based on optimal transport like the Earth Mover distance would be ideal to fix this problem. However, in practice, we ran into optimisation, training time and parameter tuning issues when using EMD. On the contrary the Chamfer loss is simple to use and fast to compute. We thus use the Chamfer loss and introduce a repulsive loss between the Gaussian means defining the probability distribution $\proba$ as a regularization: 
\begin{equation}
    \label{eq:repulse}
    \mathcal{L}_\textrm{rep}(\mathbf{\mu}) = \frac{1}{K^2} \sum_{i,j \in [0,K]} \exp({-\frac{\|\mu_i-\mu_j\|}{\sigma}}) ~,
\end{equation}
where $\sigma$ is the Gaussian standard deviations in the definition of $\proba$. We found that this simple loss lead to much more uniform distributions of points both in the 2D plane and in the reconstructed shapes.

\subsection{Joint learning on a family of shapes.}

\label{subsec:family}

We now explain how we learn jointly atlases on a collection of $N$ shapes $\shape_1, ... , \shape_N$. Since we want to share the 2D domain between the different shapes, we do not condition the probability distribution $\proba$ on the shape, and learn a single one for all shapes. On the contrary, the parametrization and chart-mappings are expected to depend on the shape. Rather than learning them completely independently for each shape, we use the auto-decoder framework~\cite{park2019deepsdf}, where we optimize for each shape $\shape_i$ feature vectors $\featurepsii$ and $\featurephii$ which we use to define respectively  $\mapTwoToSix_{\shape_i}$ and $\mapSixToTwo_{\shape_i}$. More precisely, we learn jointly for all shapes networks $\mapTwoToSix$ and $\mapSixToTwo$, and define for each shape $\shape_i$ for all $\pointgeneral\in\mathbb{R}^2$, $\mapTwoToSix_{\shape_i}(\pointgeneral)=\mapTwoToSix(\pointgeneral,\featurepsii)$ and for all $\pointgeneral \in \mathbb{R}^6$, $\mapSixToTwo_{\shape_i}(\pointgeneral)=\mapSixToTwo(\pointgeneral,\featurepsii)$. We then optimize $\mapSixToTwo$, $\mapTwoToSix$ and $\mu$ by minimizing the loss:
\begin{equation}
        \mathcal{L}_\text{full}(\mapTwoToSix , \mapSixToTwo, \mathbf{\mu},\featurepsi, \featurephi) =  \sum_{i=1}^N \mathcal{L}_\text{single}(\mapTwoToSix_{\shape_i} , \mapSixToTwo_{\shape_i}, \mathbf{\mu}) ~,
\end{equation}
where $\featurepsi=(\mathbf{f}_{\psi,1}, ... ,\mathbf{f}_{\psi,N})$ and $\featurephi=( \mathbf{f}_{\varphi,1} , ... , \mathbf{f}_{\varphi,N})$.

\paragraph{Implementation details.} 
We use latent codes $\featurephii$ and $\featurepsii$ of dimension 256. %
The architectures we use for $\mapTwoToSix$  and $\mapSixToTwo$  are MLP with 5 hidden layers of size 256 and ReLU activations. We do not use batch normalization layers. $\proba$ is defined using $K=10,000$ mixture components. %
We sample $M=10^4$ points both on the shapes and according to $\proba$ at every training iteration. We train with $\lambda_{\textrm{6D}}=1$, $\lambda_{\textrm{2D}}=10^{-2}$, $\lambda_{\textrm{cycle}}=1/M$, $\lambda_{\textrm{iso}}=10^{-4}/M$ and $\lambda_{\textrm{rep}}=1$. %

%% file: content/experiments.tex
\begin{figure}[t]
    \centering
    \includegraphics[width=\textwidth]{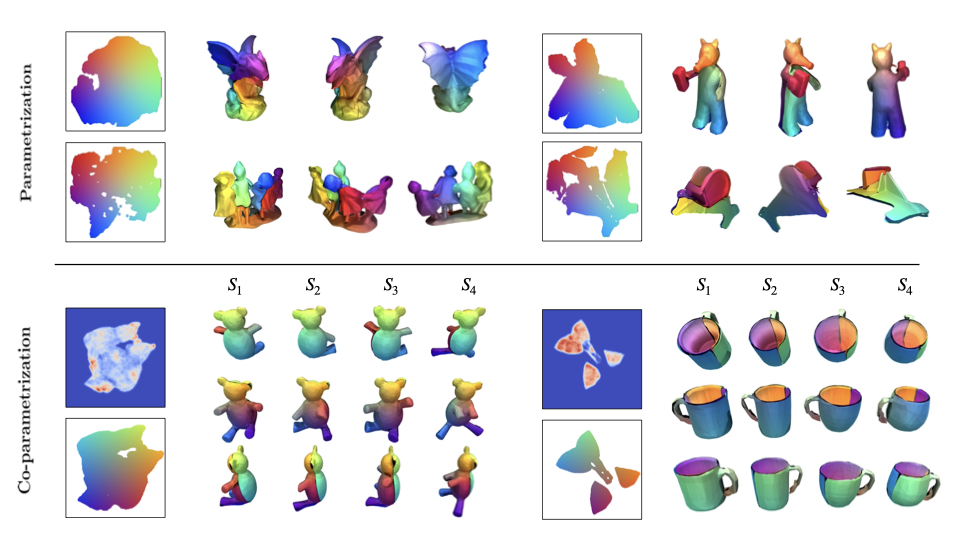}
    \caption{{\textbf{Top: single shape parametrization.} On  the shapes of~\cite{williams2019deep}, we transfer a colored mesh of the learned 2D domain (left) to 3D, which enables visualizing correspondence and cuts (right). \textbf{Bottom: co-parameterization.} For 'teddy' and 'cup' shapes of SHREC~\cite{{giorgi2007shape}}, we show the joint sampling probabilities and 2D domain, and the reconstructions of 4 shapes with 3 different viewpoints and consistent coloring. Note the quality of the parametrization compared to patch-based methods~\cite{bednarik2020shape,groueix2018}}}
    \label{fig:collection}
    \vspace{-1em}
\end{figure}

\paragraph{Datasets.} %
We use individually the shapes of Williams et al.~\cite{williams2019deep}, which come from five high-resolution scans with over a million points with associated normals. We generate the manifold meshes from this data using screened Poisson Surface Reconstruction. The resulting meshes have a variety of geometric details and different topologies, providing interesting challenges for atlas-based representations. The SHREC dataset \cite{giorgi2007shape} contains 400 manifold meshes with sparse correspondence annotations, which enables us to quantitatively evaluate the consistency of our joint atlases. For our experiments we selected categories, \textit{ant}, \textit{teddy}, \textit{cups} and \textit{armadillo}, aiming for topological and geometric diversity, and four shapes in each.

\paragraph{Reconstruction Metrics.} 
To evaluate how well our representation matches the input shape, we report two commonly used metrics: 3D Chamfer Distance and the Earth Mover's Distance (EMD). 
Given two sets of points $\mathcal{X}$ and $\mathcal{Y}$ with $M'=2000$ randomly-sampled points each, we approximate the EMD as:
\begin{equation}
    \begin{split}
        \mathcal{L}_\textrm{EMD}(\mathcal{X},\mathcal{Y})
        = \frac{1}{M'} \sum_{\mathbf{x}_i\in\mathcal{X}}
        \sum_{\mathbf{y}_j\in\mathcal{Y}} C_{i,j} \|\mathbf{x}_i-\mathbf{y}_j\|^2 ~,
    \end{split}
\end{equation}
where the association matrix $C$ is such that $C_{i,j}=1$ if point $\mathbf{x}_i$ is mapped to point $\mathbf{y}_j$ and $C_{i,j}=1$, and is computed to minimize $\mathcal{L}_\textrm{EMD}$ via Hungarian Algorithm \cite{kuhn1955hungarian}. To evaluate the reconstructions of the normals of the surfaces, we also report the distances of the normals using the associations given either by the chamfer distance or the EMD using the spatial coordinates.

\paragraph{Correspondence Metric.} \label{subsubsection:sparse}
Given a pair of shapes $\shape_1, \shape_2$ and a set of keypoints $\mathbf{p}_i \in \shape_1$, we find the corresponding points $\mathbf{q_i} \in \shape_2$ by compositing the chart-mapping and parameterization networks:
    $\mathbf{q}_i=\mapTwoToSix_{\shape_2}\circ\mapSixToTwo_{\shape_1}(\mathbf{p}_i)$.
For the baselines that do not have a chart-mapping network, we obtain correspondences for each point $\mathbf{p}_i$ using the following process: we sample points according to $\proba$, map them to $\shape_1$, select the one which image by $\mapTwoToSix_{\shape_1}$ is closest to $\mathbf{p}_i$, and map it to $\shape_2$ using $\mapTwoToSix_{\shape_2}$.
Given $M''$ ground truth correspondences $\mathbf{q}_i^\text{gt}$ for each  $\mathbf{p_i}$, we measure the L2 loss on the spatial coordinates as $\frac{1}{M''}\sum_{{i}}|\mathbf{q_i}-\mathbf{q_i^\text{gt}}|$.

\paragraph{Results and analysis. }
Qualitative results for individual shapes of Williams et al.~\cite{williams2019deep} are shown in Figure~\ref{fig:collection} (top).  Joint atlases for the SHREC dataset~\cite{giorgi2007shape},  can be seen for the 'ants' are in the teaser Figure~\ref{fig:teaser} and for 'cup' and 'teddy' in Figure~\ref{fig:collection} (bottom). Note how we manage to learn continuous 2D domains with complex topology, which we can mesh to obtain a high quality 3D mesh for the shapes. Also note that compared to the patch-based approaches we obtain few and meaningful cuts in the parametrization, without any overlap. Finally, note the consistency in the reconstruction of the different shapes, which can be visualized by consistent colors transfered from the joint 2D domain to all shapes.

\setlength{\tabcolsep}{3pt}
\begin{table}[t]
    \centering
    \resizebox{\textwidth}{!}{%

     {\tiny
     \begin{tabular}{rl|cccc|ccccc}

           \multicolumn{2}{c|}{} & 
           \multicolumn{4}{c|}{Williams et al.~\cite{williams2019deep}} & \multicolumn{5}{c}{SHREC~\cite{giorgi2007shape}} \\
           \multicolumn{2}{c|}{} & 
           \multicolumn{2}{c}{\textbf{Spatial}} & 
           \multicolumn{2}{c|}{\textbf{Normal}} &
           \multicolumn{2}{c}{\textbf{Spatial}} & 
           \multicolumn{2}{c}{\textbf{Normal}} & \textbf{Corresp.} \\
           \multicolumn{2}{c|}{} & Ch. $\downarrow$& Emd $\downarrow$& Ch. $\downarrow$& Emd $\downarrow$ & Ch. $\downarrow$ & Emd $\downarrow$ & Ch. $\downarrow$& Emd $\downarrow$ & L2 $\downarrow$ \\
          \hline
          \multicolumn{2}{l|}{\textit{Baselines}} & & & & & & & & \\
          
          $\scriptscriptstyle(1)$ & ANv1-1~\cite{groueix2018} 
          &  \cellcolor{orange}$2.3 {\scriptscriptstyle \pm0.0}$ & \cellcolor{tabred}$\boldsymbol{5.0} {\scriptscriptstyle \pm0.1}$ & $1.2 {\scriptscriptstyle \pm0.2}$ & $1.2 {\scriptscriptstyle \pm0.2}$ & $2.2{\scriptscriptstyle \selectfont \pm0.0}$ & \cellcolor{tabred}$\boldsymbol{5.4} {\scriptscriptstyle \selectfont \pm0.5}$ & $1.1 {\scriptscriptstyle \selectfont \pm0.2}$ & $1.1 {\scriptscriptstyle \selectfont \pm0.2}$ 
          & $2.6 {\scriptscriptstyle \pm0.8}$ \\ 
          
          $\scriptscriptstyle(2)$ & ANv2-1~\cite{deprelle2019learning} 
          & \cellcolor{tabred} $\boldsymbol{2.2} {\scriptscriptstyle \pm0.0}$ & $11.1 {\scriptscriptstyle \pm2.1}$ & $1.2 {\scriptscriptstyle \pm0.1}$ & $1.3 {\scriptscriptstyle \pm0.1}$ & \cellcolor{tabred}$\boldsymbol{2.0}{\scriptscriptstyle \selectfont \pm0.0}$ & $7.1 {\scriptscriptstyle \selectfont \pm1.4}$ & $1.3{\scriptscriptstyle \selectfont \pm0.2}$ & $1.3 {\scriptscriptstyle \selectfont \pm 0.2}$ 
          & $1.3 {\scriptscriptstyle \pm0.1}$ \\ 
          
          $\scriptscriptstyle(3)$ & ANv1-10~\cite{groueix2018} 
          &$2.4 {\scriptscriptstyle \pm0.0}$ & $11.4 {\scriptscriptstyle \pm0.9}$ & $1.1 {\scriptscriptstyle \pm0.0}$ & $1.1 {\scriptscriptstyle \pm0.0}$ &  $2.3{\scriptscriptstyle \selectfont \pm0.1}$ & $11.1 {\scriptscriptstyle \selectfont \pm0.4}$ & $1.1{\scriptscriptstyle \selectfont \pm0.1}$ & $1.1{\scriptscriptstyle \selectfont \pm0.1}$ 
          & $3.5 {\scriptscriptstyle \pm 1.4}$ \\
          
          $\scriptscriptstyle(4)$ & ANv2-10~\cite{deprelle2019learning} 
          & $2.2 {\scriptscriptstyle \pm0.0}$ & $11.1 {\scriptscriptstyle \pm2.1}$ & $1.2 {\scriptscriptstyle \pm0.1}$ & $1.3 {\scriptscriptstyle \pm0.1}$  & \cellcolor{orange}$2.1{\scriptscriptstyle \selectfont \pm0.0}$ & $13.4 {\scriptscriptstyle \selectfont \pm0.5}$ & \cellcolor{orange}$1.0{\scriptscriptstyle \selectfont \pm0.1}$ & $1.1 {\scriptscriptstyle \selectfont \pm 0.1} $ 
          & $2.8 {\scriptscriptstyle \pm1.6}$ \\ 
          
          $\scriptscriptstyle(5)$ & DSR-10~\cite{bednarik2020shape} 
          & $4.3 {\scriptscriptstyle \pm0.8}$ & $18.2 {\scriptscriptstyle \pm0.8}$ & $1.4 {\scriptscriptstyle \pm0.0}$ & $1.2 {\scriptscriptstyle \pm0.0}$ & $3.1 {\scriptscriptstyle \pm1.1}$ & $10.0 {\scriptscriptstyle \pm2.8}$ & $1.3 {\scriptscriptstyle \pm0.1}$ & $1.3 {\scriptscriptstyle \pm0.1} $
          & $2.6 {\scriptscriptstyle \pm1.2}$ \\
          \multicolumn{2}{l|}{\textit{Ours}} & & & & & & & & \\
          
          $\scriptscriptstyle(6)$ & $\phi$ 
          & \cellcolor{tabred} $\boldsymbol{2.2} {\scriptscriptstyle \pm0.0}$ & $15.2 {\scriptscriptstyle \pm5.9}$ & $1.0 {\scriptscriptstyle \pm0.2}$ & $1.1 {\scriptscriptstyle \pm0.2}$ & \cellcolor{tabred}$\boldsymbol{2.0}{\scriptscriptstyle \pm0.0}$ & $6.8{\scriptscriptstyle \pm0.9}$ & $1.1{\scriptscriptstyle \pm0.2}$ & $1.1{\scriptscriptstyle \pm0.2}$  
          & $2.0 {\scriptscriptstyle \pm0.8}$ \\
          
          $\scriptscriptstyle(7)$ & $\phi,n$ 
          &$2.4 {\scriptscriptstyle \pm0.1}$ & $20.3 {\scriptscriptstyle \pm3.3}$ & \cellcolor{orange}$0.3 {\scriptscriptstyle \pm0.0}$ & $0.7 {\scriptscriptstyle \pm0.0}$ & \cellcolor{orange}$2.1{\scriptscriptstyle \pm0.3}$ & $13.1{\scriptscriptstyle \pm8.0}$ & \cellcolor{tabred}$\boldsymbol{0.2}{\scriptscriptstyle \pm0.0}$ & $\cellcolor{orange}0.5{\scriptscriptstyle \pm0.0}$ 
          & $1.4 {\scriptscriptstyle \pm 1.1}$ \\
          
          $\scriptscriptstyle(8)$ & $\phi,n,r$ 
          & \cellcolor{tabred} $\boldsymbol{2.2} {\scriptscriptstyle \pm0.1}$ & $12.9 {\scriptscriptstyle \pm4.4}$ & \cellcolor{tabred}$\boldsymbol{0.2} {\scriptscriptstyle \pm0.1}$ & \cellcolor{orange}$0.6 {\scriptscriptstyle \pm0.1}$ & \cellcolor{tabred}$\boldsymbol{2.0}{\scriptscriptstyle \pm0.1}$ & $9.6{\scriptscriptstyle \pm6.6}$ & \cellcolor{tabred}$\boldsymbol{0.2}{\scriptscriptstyle \pm0.0}$ & \cellcolor{tabred}$\boldsymbol{0.4}{\scriptscriptstyle \pm0.1}$  
          & $2.8 {\scriptscriptstyle \pm0.9}$ \\ 
          
          $\scriptscriptstyle(9)$ & $\phi,n,r,p$ 
          &\cellcolor{orange}$2.3 {\scriptscriptstyle \pm0.2}$ & $13.2 {\scriptscriptstyle \pm4.0}$ & \cellcolor{orange}$0.3 {\scriptscriptstyle \pm0.1}$ & \cellcolor{orange}$0.6 {\scriptscriptstyle \pm0.1}$ & \cellcolor{tabred}$\boldsymbol{2.0}{\scriptscriptstyle \pm0.0}$ & $8.2{\scriptscriptstyle \pm2.1}$ & \cellcolor{tabred}$\boldsymbol{0.2}{\scriptscriptstyle \pm0.0}$ & \cellcolor{tabred}$\boldsymbol{0.4}{\scriptscriptstyle \pm0.0}$ 
          & $1.5 {\scriptscriptstyle \pm 1.0}$ \\

          $\scriptscriptstyle(10)$ & $\phi,\psi,n,r,p$
          & $2.6 {\scriptscriptstyle \pm0.0}$ & \cellcolor{orange}$\boldsymbol{5.1} {\scriptscriptstyle \pm0.1}$ & \cellcolor{tabred}$\boldsymbol{0.2} {\scriptscriptstyle \pm0.0}$ & \cellcolor{tabred}$\boldsymbol{0.5} {\scriptscriptstyle \pm0.0}$  & $2.2 {\scriptscriptstyle \pm0.0}$ & \cellcolor{tabred}$\boldsymbol{5.4} {\scriptscriptstyle \pm0.2}$ & \cellcolor{tabred}$\boldsymbol{0.2} {\scriptscriptstyle \pm0.0}$ & \cellcolor{tabred}$\boldsymbol{0.4} {\scriptscriptstyle \pm0.0}$ 
          & \cellcolor{orange}$1.0 {\scriptscriptstyle \pm1.0}$ \\
          
          $\scriptscriptstyle(11)$ & $\phi,\psi,n,r,p,i$ 
          & $2.6 {\scriptscriptstyle \pm0.2}$ & \cellcolor{orange} $5.5 {\scriptscriptstyle \pm1.3}$ & \cellcolor{tabred}$\boldsymbol{0.2} {\scriptscriptstyle \pm0.3}$ & \cellcolor{tabred}$\boldsymbol{0.5} {\scriptscriptstyle \pm0.2}$ & $2.3 {\scriptscriptstyle \pm0.1}$ & $ \cellcolor{orange}5.9 {\scriptscriptstyle \pm0.6}$  & \cellcolor{tabred}$\boldsymbol{0.2} {\scriptscriptstyle \pm0.0}$& \cellcolor{tabred}$\boldsymbol{0.4} {\scriptscriptstyle \pm0.0}$ 
          & \cellcolor{tabred}$\boldsymbol{0.8} {\scriptscriptstyle \pm1.0}$ \\
         \hline
     \end{tabular}}
     }
     {\tiny
     \\
     \vspace{.5em}
     \textbf{Notations} $n$: normal - $r$: repulsion loss - $p$: Probability distribution function $\pdffunction$ - $i$: Isometric regularisation of $\mappingTwoToSix$\\}
     \vspace{1em}
     \caption{\textbf{Comparison and ablation study.} We compare our method, successively adding its different componenets, to AtlasNet~\cite{groueix2018}, AtlasNetV2~\cite{deprelle2019learning} and DSR~\cite{bednarik2020shape}. We compute association between target and reconstructed 3D points using Chamfer distance and EMD, then report for each association the average distance between the points coordinates ('Spatial') and their normals ('Normals'). We also evaluate the quality of the correspondences when available. %
     All the values are scaled by a factor of $10^{-2}$. Please see text for details. }
     \vspace{-2em}
     \label{table:ablation_recons}
\end{table}

 We report quantitative results  %
in  Table~\ref{table:ablation_recons}. To account for randomness in initialization and optimization, we re-run each method four times and report mean and standard deviation.
We report results for three baselines: vanilla AtlasNet~\cite{groueix2018} (rows 1 and 3) which only learns the parameterization network, AtlasNet v2-points~\cite{deprelle2019learning} (rows 2 and 4) which learns elementary point structures in the 2D domain, and DSR~\cite{bednarik2020shape} (row 5) which uses several regularisation losses including an isometry loss. We try the first two methods with 1 and 10 UV patches. We found that methods using a single patch typically lead to similar Chamfer distances but much lower EMD. We believe that this is because using a single patch encourages points to be more uniformly distributed on the surface. We also found that DSR~\cite{bednarik2020shape} provided quantitativelty slightly worse results, which is consistent with what was reported in this paper that mainly aims at visual quality. %
 
\begin{figure}[t]
\centering
    \includegraphics[height=2.3cm]{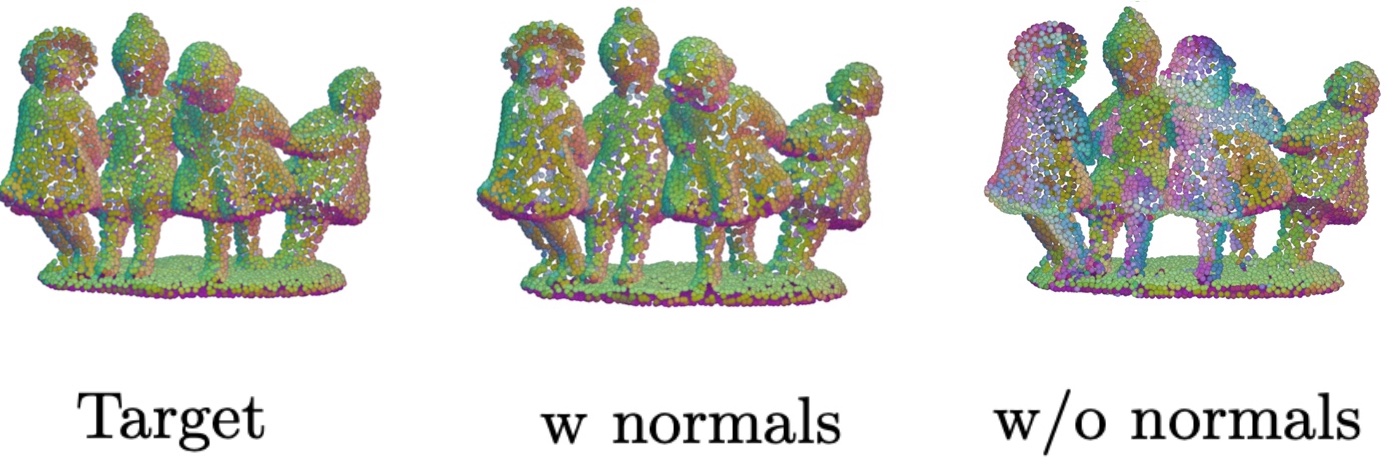}~~~~~~\vline~~~~~~
    \includegraphics[height=2.3cm]{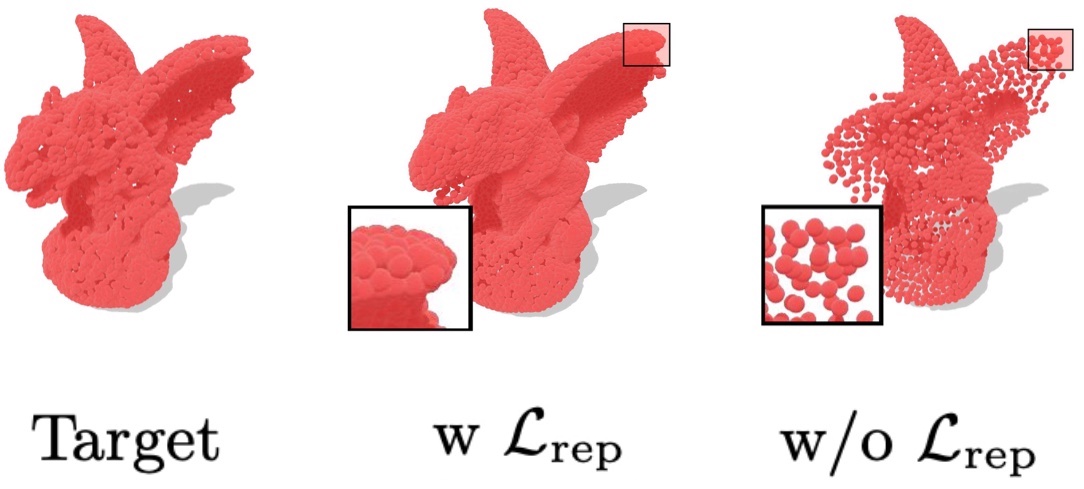}
     \caption{\textbf{Impact of the normals and the repulsion loss.} Note the variation of the color-coded normal directions (left) and the variations in the point density (right). %
     } 
     \label{fig:reconstruction}
\end{figure}

 We evaluate our method by successively adding our different components. We start (row 6) by only using our parametrization network, learning a fixed set of input point positions with only 3D Chamfer distance as supervision, similar to AtlasNet v2 with 1 patch and obtained similar results. \\
We then add normals to the Chamfer loss (row 7), which unsurprisingly gives a very strong boost to the normal metrics. Qualitatively, the effect can be seen in the left part of Figure~\ref{fig:reconstruction} where the orientation of the normals is color-coded for each point: without the normals in the loss, a significant part of the normals are back-facing. Adding the normals in the loss however has a second undesired effect and significantly increases the EMD. This can be understood qualitatively by looking at the results, where we see greatly varying density of points over the reconstructed shapes and in the 2D space. We interpret this as the fact that adding normals in the loss complexifies the loss landscape and adds bad local minima. \\
To avoid this effect, we add our repulsive regularization to the loss (row 8). We can see this improves the EMD results without degrading the Chamfer results and normal consistency. The effect on the points density is also striking qualitatively, as visualized in the right part of Figure~\ref{fig:reconstruction}.\\
At this point, the parametrization is still only optimized on a set of points. To recover an interpretation of the parametrization as a continuous mapping of a 2D domain, we introduce our sampling probability distribution (row 9) which has little quantitative effect but is crucial to define a continuous 2D mapping. \\
We can then introduce our chart-mapping, together with the 2D reconstruction and cycle losses (row 10). It can be seen that beside its theoretical interest and benefits, it has a clear quantitative impact: both the EMD and the correspondence metrics are clearly boosted.\\
Finally, adding the isometric regularization (row 11) provides a small additional boost in term of correspondence quality, i.e., consistency between different jointly learned atlases.

\begin{figure}[t]
    \scriptsize
    \captionsetup[subfigure]{labelformat=empty}
     \centering
    \begin{subfigure}[b]{.16\textwidth}
         \centering
         \includegraphics[width=\textwidth, trim={20cm 5cm 20cm 3cm},clip ]{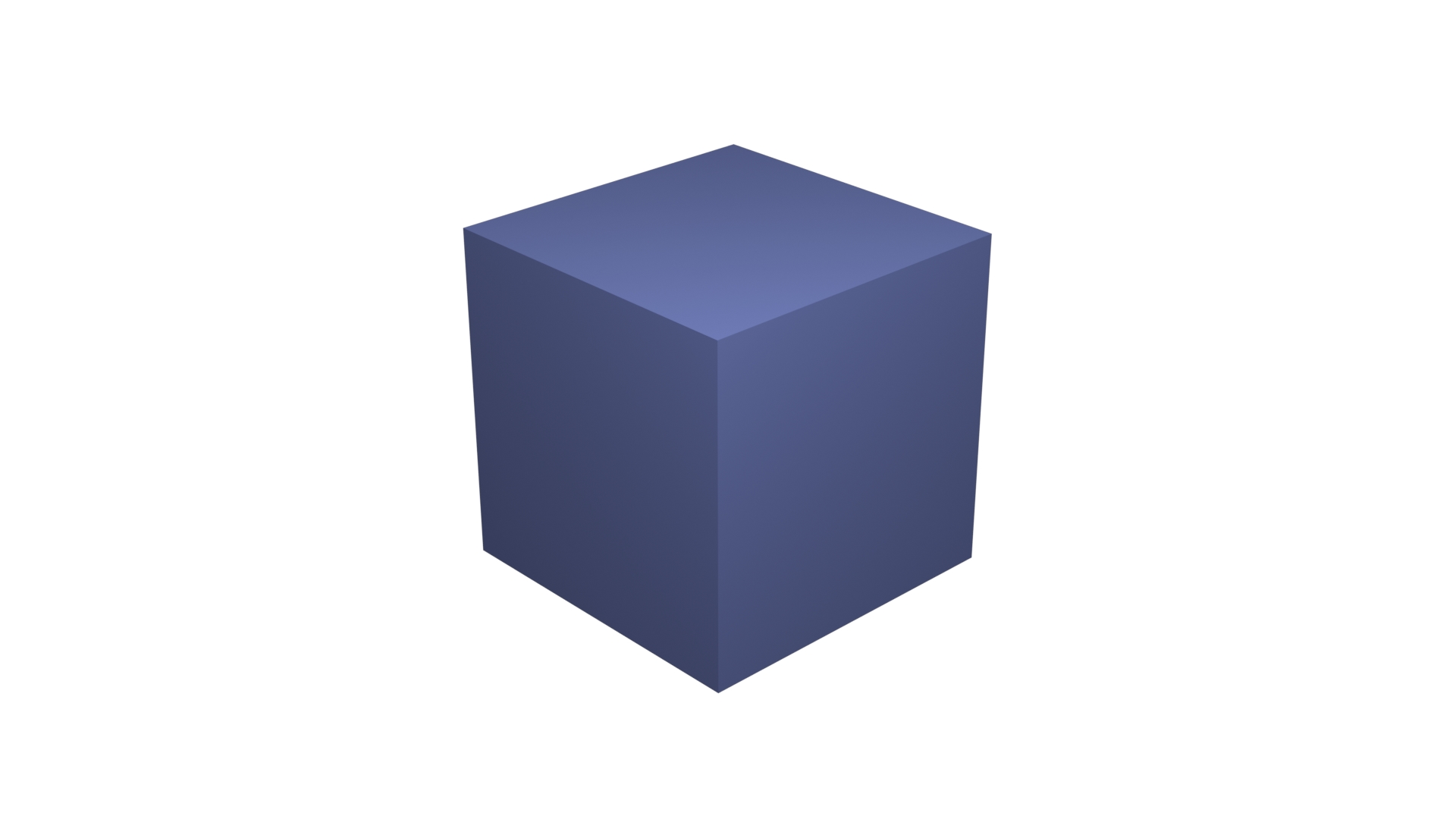}
         \caption{Target}
     \end{subfigure}~~~
     \begin{subfigure}[b]{.16\textwidth}
         \centering
         \includegraphics[width=\textwidth, ,clip ]{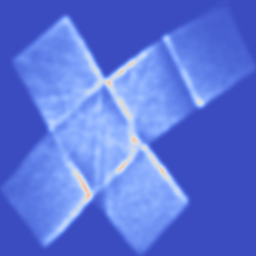}
         \caption{ours}
     \end{subfigure}~~~
     \begin{subfigure}[b]{.16\textwidth}
         \centering
         \includegraphics[width=\textwidth, ,clip ]{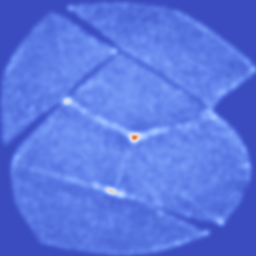}
         \caption{w/o iso}
     \end{subfigure}~~~
      \begin{subfigure}[b]{.16\textwidth}
         \centering
         \includegraphics[width=\textwidth, ,clip]{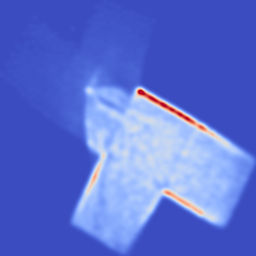}
         \caption{w/o normals}
     \end{subfigure}~~~
     \begin{subfigure}[b]{.16\textwidth}
         \centering
         \includegraphics[width=\textwidth, ,clip]{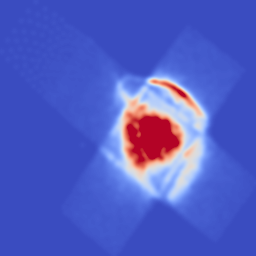}
         \caption{w/o rep.}
     \end{subfigure}
     \hfill
     \caption{\textbf{Qualitative ablation results.} Given a target cube shape, we show the probability distribution $\proba$ learned by our method and three key ablations: not using the isometric loss, not using the normals, and not using the repulsive loss.  %
     }
     \label{fig:repulsion}
     \label{fig:qualitative_ablation}
     \vspace{-1em}
\end{figure}

Note that all of our regularization losses remain necessary to the success of our method, as can be visualized qualitatively in Figure~\ref{fig:qualitative_ablation}, where we use our architecture to learn the reconstruction of a simple cube and visualize the learned 2D sampling distribution. Without the normal loss, we observe a concentration of the density on edges, an equivalent of point collapse described in~\cite{bednarik2020shape}. Without the repulsion loss, the density is concentrated on a single face. Without the isometry loss, the shape of the cube is highly distorded. FWith our full method the net of the cube is clear and the point density relatively uniform.

%% file: content/conclusion.tex
We propose a novel technique for representing a shape or a collection of shapes, with two key differences with respect to prior work on atlas-like representations. First, we joinlty learn two maps, a parameterization and a chart-mapping. Second, we learn the 2D domain on which the parametrization is defined. This makes our representation much closer than previous works to be a proper atlas, i.e., one that defines an homeomorphism between 3D shape and a 2D domain. It also enables co-parameterization, where homeomorphisms are learned between several shapes and a single 2D domain, which can have applications for consistent texturing and correspondence estimation. We further offer  other technical contributions, such as learning the 2D domain by optimizing a sampling probability distribution, analyzing the effect of incorporating normals in the optimization, and introducing a repulsive loss to have a more even point distribution. %
We believe our work is an important step towards learning consistent atlases and it will inspire future work on further improving quality of atlases, such as modeling transition maps, minimizing seams, and distortion.